\documentclass[10pt,conference]{IEEEtran}

\usepackage[dvipsnames]{xcolor}
\usepackage{color}

\definecolor{QCPbeige}{HTML}{E7D5AB} 
\definecolor{QCPblue}{HTML}{479093} 
\definecolor{QCPdarkblue}{HTML}{2C6071} 
\definecolor{QCPgray}{HTML}{333743} 
\definecolor{QCPlightblue}{HTML}{A6CBB9} 
\definecolor{QCPred}{HTML}{E08B8A} 

\usepackage{pgfplots}
\pgfplotsset{compat = newest}
\usepackage{adjustbox}
\usepackage[utf8]{inputenc}
\usepackage{physics}
\usepackage{bm} 
\usepackage{braket}
\usepackage{dsfont}
\usepackage{amsmath}
\usepackage{amssymb}
\usepackage{amsthm}
\usepackage{thmtools} 
\usepackage{mathtools}
\usepackage{tikz}
\usetikzlibrary{shapes.geometric,backgrounds,positioning,shapes.geometric,decorations.markings,decorations.pathreplacing,arrows,knots,hobby,angles,quotes}
\tikzset{meter/.append style={draw, inner sep=5, rectangle, font=\vphantom{A}, minimum width=20, line width=.4,
 path picture={\draw[black] ([shift={(.1,.2)}]path picture bounding box.south west) to[bend left=50] ([shift={(-.1,.2)}]path picture bounding box.south east);\draw[black,-latex] ([shift={(0,.1)}]path picture bounding box.south) -- ([shift={(.3,-.1)}]path picture bounding box.north);}}} 
\makeatletter
\newcommand{\gettikzxy}[3]{%
  \tikz@scan@one@point\pgfutil@firstofone#1\relax
  \edef#2{\the\pgf@x}%
  \edef#3{\the\pgf@y}%
}
\makeatother
\usepackage{graphics}
\usepackage{float}
\usepackage{arc-blochsphere}

\usepackage[colorlinks]{hyperref}
\hypersetup{
    colorlinks  = true,
    citecolor   = PineGreen,
    linkcolor   = PineGreen,
    urlcolor    = PineGreen
}

\usepackage{etoolbox}
\apptocmd{\sloppy}{\hbadness 10000\relax}{}{}

\usepackage{silence}
\usepackage{afterpage} 

\makeatletter
\def\ps@IEEEtitlepagestyle{%
  \def\@oddfoot{\mycopyrightnotice}%
  \def\@evenfoot{}%
}
\def\mycopyrightnotice{%
  {\footnotesize
  \hfill 
  \parbox{\textwidth}{%
  © 2025 IEEE.  Personal use of this material is permitted.  Permission from IEEE must be obtained for all other uses, in any current or future media, including reprinting/republishing this material for advertising or promotional purposes, creating new collective works, for resale or redistribution to servers or lists, or reuse of any copyrighted component of this work in other works.}
  \hfill}
}
\makeatother

        \DeclareMathOperator{\ancilla}{a} 
        
        \DeclareMathOperator{\opt}{opt} 
        
        \newcommand*{\R}{\mathbb{R}}
        \newcommand*{\C}{\mathbb{C}}

        \newcommand*{\hil}{\mathcal{H}}
        
            \newcommand*{\anchil}{\mathcal{H}_{\ancilla}} 
            \newcommand*{\one}{\mathds{1}} 
            \newcommand*{\malpha}{M_{\bm{\alpha}}} 

\begin{document}

\title{From barren plateaus through fertile valleys: Conic extensions of parameterised quantum circuits}

\author{\IEEEauthorblockN{Lennart Binkowski\IEEEauthorrefmark{1}, Gereon Koßmann\IEEEauthorrefmark{2}, Tobias J. Osborne\IEEEauthorrefmark{1}, René Schwonnek\IEEEauthorrefmark{1}, Timo Ziegler\IEEEauthorrefmark{1}${}^{,}$\IEEEauthorrefmark{3}}
\IEEEauthorblockA{\IEEEauthorrefmark{1}Institut f\"ur Theoretische Physik\\
Leibniz Universit\"at Hannover, Hannover, Germany}
\IEEEauthorblockA{\IEEEauthorrefmark{2}Institute for Quantum Information\\
RWTH Aachen University, Aachen, Germany}
\IEEEauthorblockA{\IEEEauthorrefmark{3}Email: timo.ziegler@itp.uni-hannover.de}}

\maketitle
\thispagestyle{IEEEtitlepagestyle}

\begin{abstract}
    Optimisation via parameterised quantum circuits is the prevalent technique of near-term quantum algorithms. However, the omnipresent phenomenon of barren plateaus -- parameter regions with vanishing gradients -- sets a persistent hurdle that drastically diminishes its success in practice.

    In this work, we introduce an approach -- based on non-unitary operations -- that favours jumps out of a barren plateau into a fertile valley. These operations are constructed from conic extensions of parameterised unitary quantum circuits, relying on mid-circuit measurements and a small ancilla system. We further reduce the problem of finding optimal jump directions to a low-dimensional generalised eigenvalue problem.
    
    As a proof of concept we incorporate jumps within state-of-the-art implementations of the Quantum Approximate Optimisation Algorithm (QAOA).
    We demonstrate the extensions' effectiveness on QAOA through extensive simulations, showcasing robustness against barren plateaus and highly improved sampling probabilities of optimal solutions.
\end{abstract}


\section{Introduction}\label{section:Introduction}

Optimisation problems play a pivotal role throughout the natural sciences, society, and beyond.
Decades of development have led to highly optimised specialised algorithms for the solution of optimisation problems, particularly those with a convex structure \cite{Boyd2004ConvexOptimization}.
Most prominent here is the simplex algorithm \cite{Dantzig1990OriginsOfTheSimplexMethod} which continues to be the method of choice for numerous commercially available solvers.
Other examples include the least squares method and linear programming, tools of such fundamental utility, they are now ubiquitous.
The solution of large optimisation problems is now often bottlenecked by the runtime or available memory and improvements by even constant factors can be of considerable industrial significance.

\begin{figure*}[!ht]
\begin{minipage}[t]{0.7\linewidth}
\includegraphics[width=.9\linewidth, trim={2cm, 16.5cm, 4cm, 8cm}]{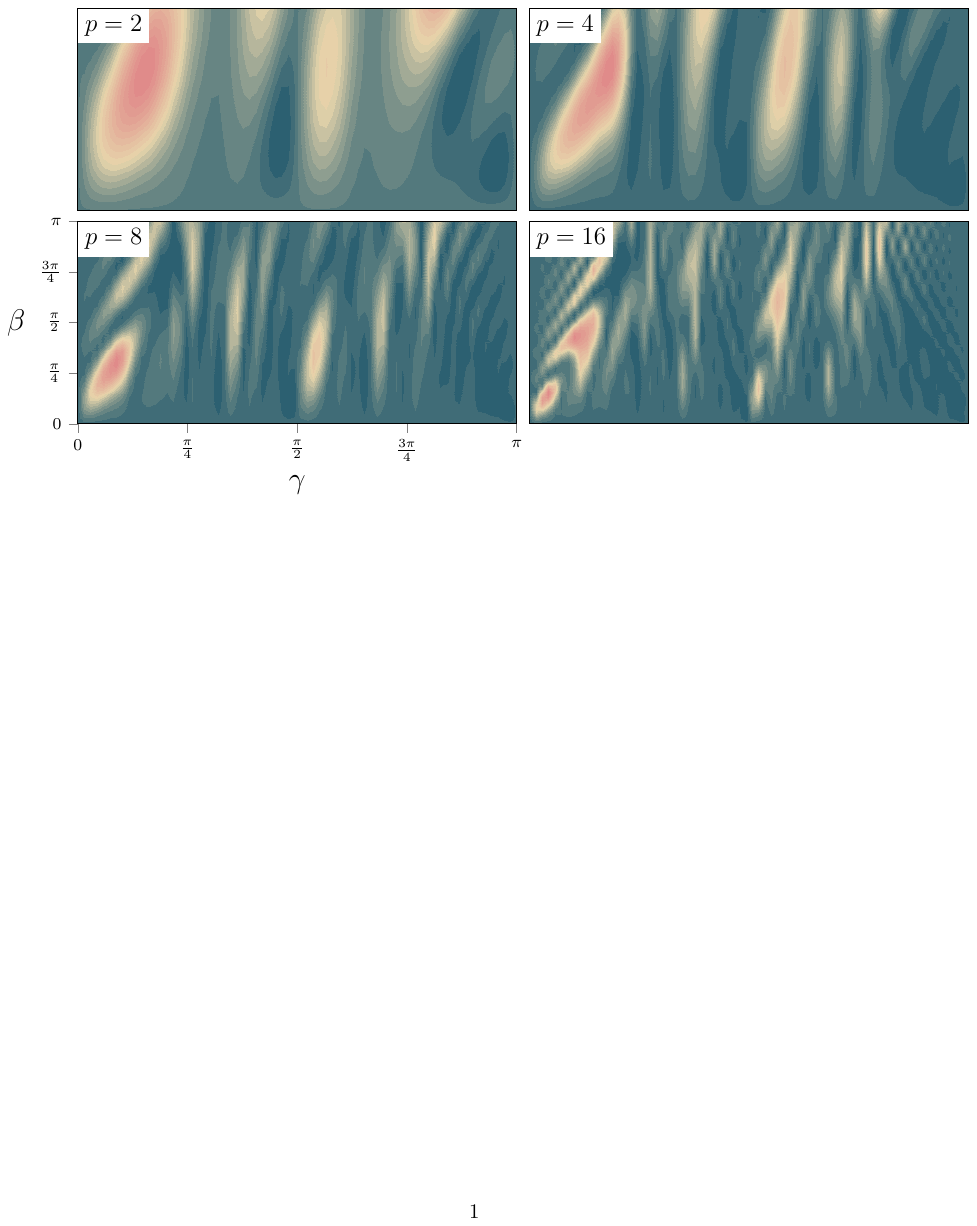}
\end{minipage}\hfill%
\begin{minipage}[t]{0.27\linewidth}
\includegraphics[trim={0 -.7cm 0 0}]{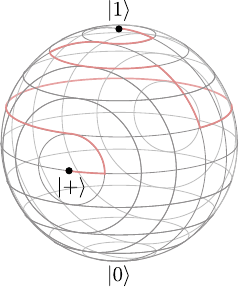}
\end{minipage}
\caption{\textbf{a)} QAOA optimisation landscapes.
The optimisation landscape is illustrated for a QAOA$_{p}$ PQC with $p =2, 4, 8$, and $16$ as applied to a hamiltonian instance with a single marked state $\ket{0001}$.
For the sake of illustration the optimisation parameters $\beta_{j}$ and $\gamma_{j}$ have been set equal, i.e., $\beta_{j}=\beta$ (the $x$ axis) and $\gamma_{j} = \gamma$ (the $y$ axis).
This simplification nevertheless captures the essence of the barren plateau phenomenon:
One observes that the optimal value of the variational parameters occurs in a tiny region in the bottom left of the figure.
The rest of the landscape is comparatively flat.
\textbf{b)} An exemplary QAOA$_{4}$ run on a single qubit, i.e., the Bloch sphere.
The direct path between the initial state $\ket{+}$ and the target state $\ket{1}$ runs through the Bloch sphere and is therefore not accessible via the unitary QAOA gates.
Instead, their application traces a complicated path on the surface of the Bloch sphere.
}\label{figure:Challenge}
\end{figure*}

Quantum computers are now poised to bring about a revolution in the computational sciences with a host of new approaches to solve optimisation problems.
Quantum advantage for a variety of use cases is now considered very likely in the long term, once plentiful and cheap \emph{logical} qubits are available.
However, in the near and medium term the situation is far from clear;
in the coming years quantum information processing devices will continue to operate above the fault-tolerance threshold, and hence directly suffer from decoherence \cite{Preskill2018QuantumComputingInTheNISQEraAndBeyond}.
Thus it is of paramount importance to determine whether noisy intermediate scale quantum (NISQ) devices are capable of supplying a meaningful accelerated solution for an optimisation problem of practical relevance, a task which has been investigated in earnest during the past decade. 

The currently most prominent NISQ-compatible approach for the quantum-accelerated solution of optimisation problems is furnished by \emph{variational quantum algorithms} (VQA) applied to \emph{parameterised quantum circuits} (PQC) \cite{Cerezo2021VariationalQuantumAlgorithms}.
Prominent here is the \emph{quantum approximate optimisation algorithm} (QAOA) \cite{Farhi2014AQuantumApproximateOptimizationAlgorithm}.
This ansatz has received extraordinary interest since its appearance in 2014, and has been generalised in many directions, most notably via the quantum alternating operator ansatz \cite{Hadfield2019FromTheQuantumApproximateOptimizationAlgorithmToAQuantumAlternatingOperatorAnsatz}.
Despite a decade of advances there has been so far no conclusive demonstration of a quantum speedup for a problem of industrial relevance.\footnote{
    There is a significant debate about the criteria to determine what would constitute a quantum speedup.
    We adopt the stronger interpretation that a quantum speedup has been realised when a quantum device has supplied a solution to an industrially significant problem which is either of better quality than that supplied by the best-available classical solvers, or which was obtained using fewer resources that the best available classical solvers.
    By ``resources'' we mean here either wall time, energy usage, i.e., effectively money.
}
This is due, in part, to the fact that presently available quantum devices are still small and noisy, so they are often easily simulated by classical methods. However, there are more fundamental reasons why speedups from variational quantum heuristics are subtle and nuanced. 

A crucial challenge facing variational quantum heuristics is that the domain where they can offer a speedup compared with state-of-the-art classical methods such as dynamic programming, branch-and-bound, and SAT solvers, is highly compressed and restricted to problems with additional structure.
A case in point here is the realisation that the QAOA with a constant number of layers for the archetypal MAXCUT problem is outperformed, even in principle, by classical solvers for a variety of problem instances \cite{Guerreschi2019QAOAForMaxCutRequiresHundredsOfQubitsForQuantumSpeedUp,Barak2022ClassicalAlgorithmsAndQuantumLimitationsForMaximumCutOnHighGirthGraphs}.
In order to outperform a classical solver it becomes necessary to exploit a parameterised quantum circuit ansatz which is comprised of many layers. Unfortunately such \emph{deep} PQCs suffer from a new problem, namely the \emph{barren plateau} phenomenon, leading to vanishing gradients \cite{McClean2018BarrenPlateausInQuantumNeuralNetworkTrainingLandscapes,Cerezo2021CostFunctionDependentBarrenPlateausInShallowParametrizedQuantumCircuits,Campos2021AbruptTransitionsInVariationalQuantumCircuitTraining}. 
Interestingly, the barren plateau phenomena is not a significant problem when studying similar models via classical tensor network algorithms such as the density matrix renormalisation group (DMRG) \cite{Schollwock2005TheDensityMatrixRenormalizationGroup,Schollwock2011TheDensityMatrixRenormalizationGroupInTheAgeOfMatrixProductStates,Bridgeman2017HandwavingAndInterpretiveDanceAnIntroductoryCourseOnTensorNetworks}. This is no accident and we exploit crucial lessons learnt in the development of the DMRG in our constructions.

In this paper we take aim at the barren plateau phenomenon and describe a general-purpose method -- applicable to a majority of parameterised quantum circuit families -- to directly overcome vanishing gradients.
We exploit several key innovations:
(1) we describe how to extend, in a natural and systematic way, a PQC variational class to include non-unitary gates so that the PQC ansatz becomes a variational \emph{cone};
(2) we then explain how a barren plateau may be avoided by taking a step \emph{through} the variational cone, making use of efficient and easily measurable moment matrices;
(3) the requisite non-unitary gate is implemented using the linear combination of unitaries (LCU) method \cite{Chakraborty2024ImplementingAnyLinearCombinationOfUnitariesOnIntermediateTermQuantumComputers,Childs2012HamiltonianSimulationUsingLinearCombinationsOfUnitaryOperations}.
Finally we numerically demonstrate, on the basis of direct simulations for challenge benchmarks, that the resulting optimisation method outperforms the corresponding variational quantum algorithm.
We exploit several crucial techniques to achieve this goal, in particular, the variational cone is achieved generalising the ansatz introduced in \cite{Bharti2022NoisyIntermediateScaleQuantumAlgorithmsForSemidefiniteProgramming}.
The use of moment matrices to estimate the derivative then exploits ideas present in  \cite{McClean2017HybridQuantumClassicalHierarchyForMitigationOfDecoherenceAndDeterminationOfExictedStates}. The LCU method we exploit was introduced in \cite{GuiLu2006GeneralQuantumInterferencePrincipleAndDualityComputer}.
The optimisation method via generalised eigenvalue problems originates, e.g., in the study of the DMRG via matrix product states \cite{Schollwock2005TheDensityMatrixRenormalizationGroup,Wilson1975TheRenormalizationGroupCriticalPhenomenaAndTheKondoProblem} (similar ideas were also recently considered in \cite{Huang2019NearTermQuantumAlgorithmsForLinearSystemsOfEquations} in the context of linear equations). 
The ansatz pursued in this paper appears to be the tip of the iceberg of a deeper and general framework which we christen \emph{quantum conic programming} (QCP).

\section{Preliminaries}\label{section:Preliminaries}

\begin{figure*}[!ht]
    \begin{minipage}[t]{0.27\linewidth}
        \includegraphics{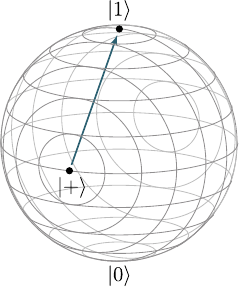}
    \end{minipage}\hfill
    \begin{minipage}[t]{0.7\linewidth}
        \includegraphics{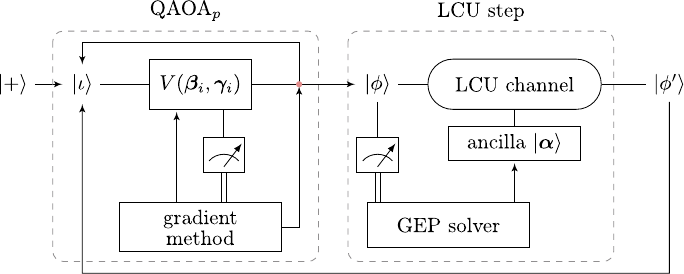}
    \end{minipage}
\caption{\textbf{a)} Step through the Bloch sphere. For the single qubit example, the direct (and therefore optimal) path between the initial state $\ket{+}$ and target state $\ket{1}$ is depicted.
Rather than traversing the Bloch sphere's surface, it directly goes through its interior.
This is the simplest example of an optimal search direction achievable by an LCU step, but not by any unitary PQC.
\textbf{b)} LCU-assisted QAOA. First, we initialise the main register in the state $\ket{\iota} = \ket{+}$ and classically optimise the circuit parameters $\bm{\beta}$, $\bm{\gamma}$ based on a gradient method.
In each subsequent iteration, we reset the system to the initial state $\ket{\iota}$ and input the parameter values of the previous round into the classical optimiser.
In this manner, we iteratively improve the parameters until a vanishing gradient is registered.
Then the current output state $\ket{\phi}$ is instead processed by the LCU step.
Here, we first measure the moment matrices $\mathbf{E}$ and $\mathbf{H}$ via Hadamard tests.
The obtained three-dimensional GEP $\mathbf{H} \bm{\alpha} = \lambda \mathbf{E} \bm{\alpha}$ is solved by a classical GEP solver and the state $\ket{\bm{\alpha}}$, prepared in the ancilla register, serves to implement the LCU channel that updates $\ket{\phi} \to \ket{\phi'}$.
In an iterative scheme, the latter state may now serve as the initial state $\ket{\iota}$ of another iteration of QAOA$_{p}$.
}\label{figure:Method}
\end{figure*}

We consider generic (unconstrained) combinatorial optimisation problems (COP), where the goal is to minimise an $n$-bit objective function $f$.
Such problems cover a vast range of applications throughout the natural sciences \cite{Garey1990ComputersAndIntractabilityAGuideToTheTheoryOfNPCompleteness}.
The standard procedure to formulate such a classical problem for study via a quantum computer is as follows.
First each bit string $\mathbf{z} = z_1z_2\cdots z_n$ is identified with a computational basis state $\ket{\bm{z}}$ of the $n$-qubit hilbert space $\hil = \C^{2^{n}}$.
The classical objective function is then encoded into a diagonal hamiltonian $H\ket{\bm{z}} = f(\bm{z}) \ket{\bm{z}}$.
In this way $\bm{z}$ is an optimal solution to the COP if and only if $\ket{\bm{z}}$ is a ground state of $H$. 

To variationally approximate such a ground state with a quantum computer a parameterised manifold of easily preparable states $\ket{\bm{\theta}} = V(\bm{\theta}) \ket{\iota}$ is introduced by specifying an initial state $\ket{\iota}$ and a unitary PQC $V(\bm{\theta}) \equiv V_L(\theta_L) \cdots V_2(\theta_2)V_1(\theta_1)$.
It is assumed that the PQC is comprised of a product of $L$ easy to implement quantum gates $V_j(\theta_j)$, each of which depends on a parameter $\theta_j$.
The energy expectation value $E(\bm{\theta}) \equiv \braket{\bm{\theta} | H | \bm{\theta}}$ is then empirically estimated via measurement on the quantum computer;
and the empiric average $\hat{E}(\bm{\theta})$ is passed to a classical optimiser which updates the parameter values $\bm{\theta} \mapsto \bm{\theta} + \delta \bm{\theta}$.
The above steps are then repeated with the updated parameter values.
After some termination condition is fulfilled, e.g.\ some threshold for the parameter updates is surpassed, the last prepared ansatz state $\ket{\bm{\theta}}$ is returned.
For example, the QAOA$_{p}$ is initialised within the uniform superposition $\ket{+}$ of all computational basis states and is built from the alternating $p$-fold application of \emph{mixer} gates $\exp(- i \beta B)$, $B = \sum_{k = 1}^{n} \sigma_{x}^{(k)}$, and \emph{phase separator} gates $\exp(- i \gamma H)$ as PQC.

The variational minimisation of $E(\bm{\theta})$ suffers, in general, from a proliferation of local minima, as illustrated in \hyperref[figure:Challenge]{Figure 1 \textbf{a)}}.\footnote{
    These minima result from saddle points on the parameterised manifolds by effectively restricting the available search directions \cite{Kossmann2022DeepCircuitQAOA}.
}
And even for a favourable optimisation landscape, the actual phase space trajectory can be unnecessarily complicated as depicted on the Bloch sphere (single qubit) in \hyperref[figure:Challenge]{Figure 1 \textbf{b)}}.
Further, if one unwisely initialises the parameters $\bm{\theta}$ with random values then the gradient of $E(\bm{\theta})$ is typically exponentially small (in the total number of qubits);
a phenomenon known as barren plateau.
Such vanishing gradient problems severely hinder the application of VQAs to general COPs.
The origin of these phenomena is easy to explain.
The optimal direction $\ket{\delta \psi}$ to update a given state $\ket{\psi}$ in order to minimise the energy $\langle\psi|H|\psi\rangle$ is given by the vector pointing in the direction of decreasing energy.
However, the variational manifold, embedded in the high-dimensional hilbert space $\mathcal{H}$, is a typically non-trivially curved submanifold of much lower dimension.
Thus, generically, the vector $\ket{\delta \psi}$ is almost orthogonal to the tangent space of the variational manifold, so there is no available direction within the manifold that appreciably reduces the energy.
Ideally, when the gradient becomes small, one would like to be able to take a step ``outside'' of the manifold.
One of our main contributions is a systematic method to do exactly this.

\section{Methods}\label{section:Methods}

\begin{figure*}[!th]
    \begin{minipage}[t]{0.45\linewidth}
        \includegraphics[scale=0.402]{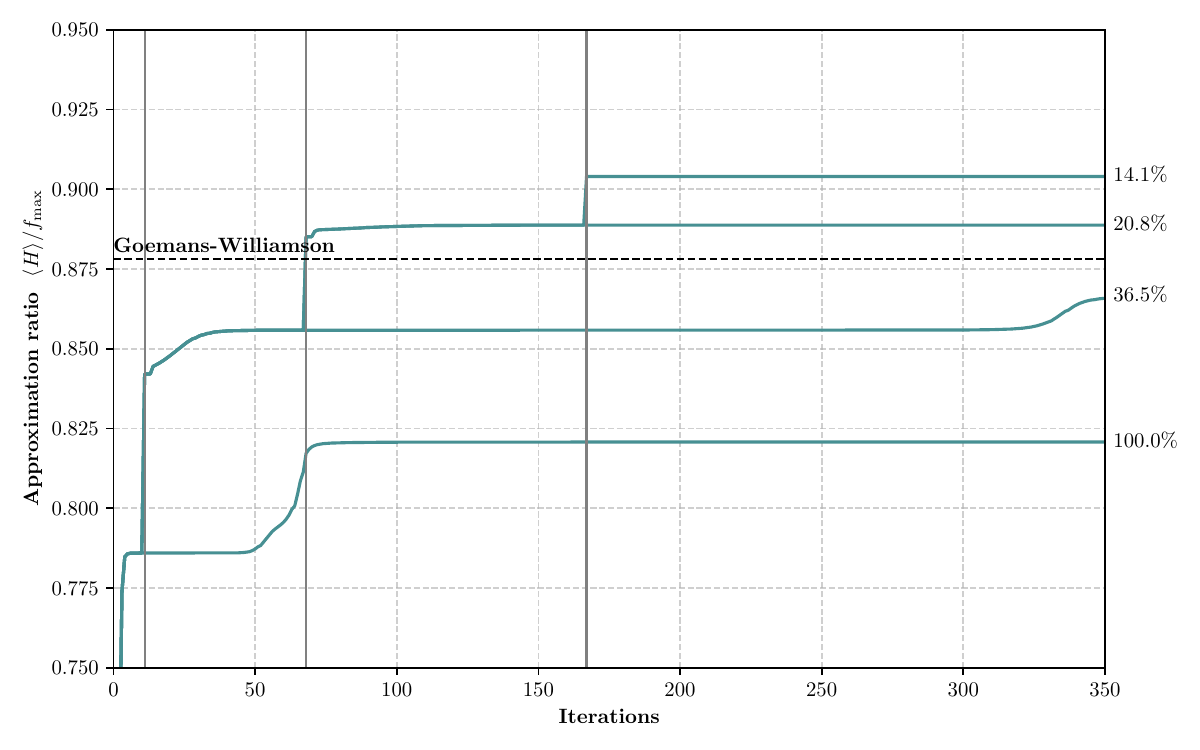}
    \end{minipage}\hfill
    \begin{minipage}[t]{0.55\linewidth}
        \includegraphics[scale=0.605]{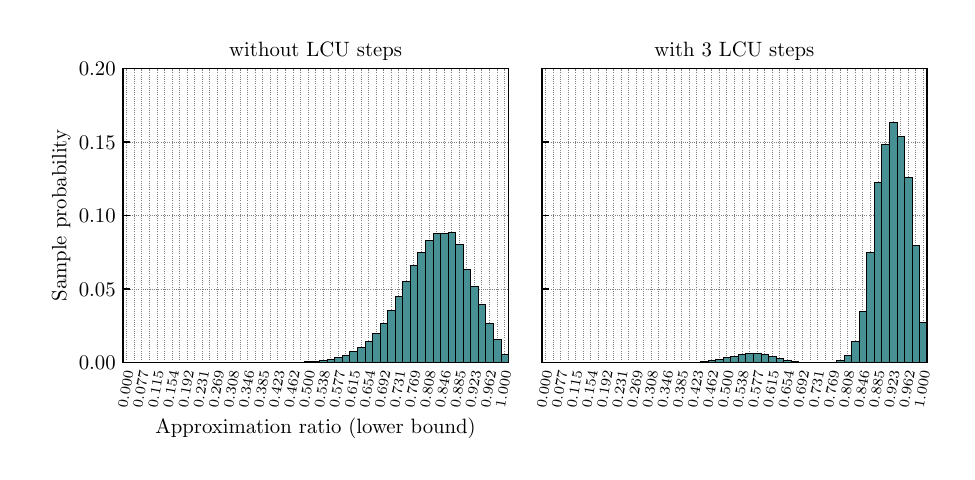}
    \end{minipage}
\caption{
Performance of the QAOA with and without LCU assistance.
We simulate the QAOA with a straight-forward gradient descent as classical optimisation routine on a 22-qubit instance of MAXCUT defined on a 3-regular graph.
\textbf{a)} The green lines show the optimisation quality -- quantified as the ratio between the expectation value $\braket{H}$ and the optimal value $f_{\max}$ -- for both the unassisted and the LCU-assisted QAOA across iterations.
Black vertical lines mark the three points at which LCU steps are applied.
The unassisted QAOA$_{2}$ stagnates already after roughly 10 iterations, reaching a plateau at a ratio of $0.785$.
In comparison, a single LCU step improves the approximation ratio to $0.842$, albeit with a reduced success probability of $36.5\%$.
Each newly started QAOA$_{2}$ branch evolves from the stagnated output states of previous QAOA branches, effectively re-optimizing from a higher-depth ansatz while keeping all prior parameters fixed.
This deeper ansatz helps overcome barren plateaus (but not local optima) by introducing new variational degrees of freedom, but each QAOA$_{2}$ branch still quickly reaches its own plateau and contributes only marginal further improvement.
Crucially, it is the LCU steps that enable the algorithm to escape local optima.
Successive applications of LCU steps continue to raise the approximation ratio significantly.
Notably, after two LCU steps, the performance surpasses that of the Goemans-Williamson algorithm, and a third LCU step elevates the approximation ratio well beyond $0.9$.
\textbf{b)}
Histograms for cumulative probabilities by approximation ratio. 
Computational basis states are grouped into bins according to their individual approximation ratios, with each bin covering the interval between two consecutive $x$-axis labels. The total sample probability of all basis states within each bin is then summed.
The left panel shows the distribution without any LCU assistance, where the output state has minimal support on high quality solutions (approximation ratio $> 0.9$).
In contrast, after three LCU steps (right), the output state becomes sharply concentrated around basis states with approximation ratios near $0.9$, indicating a substantial shift in probability toward high-quality solutions.
}
\label{figure:Results}
\end{figure*}

The core idea of our method is to start an optimisation with a usual unitary PQC and execute a parameterised non-unitary gate $\mathcal{M}_{\bm{\alpha}}$ as soon as the unitary PQC-based method stagnates.
One then appends and optimises another unitary PQC.
Once this optimisation stagnates again the next non-unitary gate is employed, and so on (see \hyperref[figure:Method]{Figure 2 \textbf{b)}}).

For such a method's viability, it is essential to consider a class of non-unitary gates $\mathcal{M}_{\bm{\alpha}}$ that are efficiently implementable, on one hand, and admit an efficient optimisation of the parameters $\bm{\alpha}$, on the other.

As we will see in the following, linear combinations of (appropriately chosen) unitaries (LCUs) do match these criteria.
Henceforth, we will consider a parameterised ansatz class for non-unitary gates $\mathcal{M}_{\bm{\alpha}}$ that act as
\begin{align}\label{equation:AnsatzClass}
    \mathcal{M}_{\bm{\alpha}} \ket{\phi} = \frac{\malpha \ket{\phi}}{\norm{\malpha \ket{\phi}}},\quad \malpha = \sum_{i = 1}^{\ell} \alpha^{\vphantom{\dag}}_{i} U^{\vphantom{\dag}}_{i}
\end{align}
where the $U_{i}$ are fixed unitary quantum gates and $\alpha_{i}$ are complex parameters.
The unitaries $U_{i}$ are selected from parameter-independent components of the preceding PQC.
In the concrete example of the QAOA$_{p}$, we choose $U_{1} = \exp(- i \delta_{1} B)$ and $U_{2} = \exp(- i \delta_{2} H)$ with fixed angles $\delta_{1}$ and $\delta_{2}$, and append $U_{3} = \one$.
By including the identity gate, the ansatz class is always able to reproduce the initial guess $\ket{\phi}$.
Hence, minimising over the ansatz class cannot decrease the overall optimisation quality.

Further, gates of the form \eqref{equation:AnsatzClass} can produce states that are not restricted to the initial parameterised manifold.
Therefore, if the optimal direction is orthogonal to the manifold's tangent space, the above ansatz class still has the chance to capture it.

The computation of optimal parameters $\bm{\alpha}$ translates to finding the minimisers for the optimisation:
\begin{align}\label{equation:ParameterOptimisation}
\begin{split}
    & \min_{\bm{\alpha} \in \C^{\ell}} \braket{\phi | \malpha^{\dag} H \malpha^{\vphantom{\dag}} | \phi} \\
    & \text{s.t. } \braket{\phi | \malpha^{\dag} \malpha^{\vphantom{\dag}} | \phi} = 1
 \end{split}
\end{align}
The emergence of a variational cone can be best understood in the Heisenberg picture:
By identifying the parameter values $\bm{\alpha}$ with the observables $\malpha^{\dag} H \malpha^{\vphantom{\dag}}$, \eqref{equation:ParameterOptimisation} can also be viewed as an optimisation task within the cone of positive observables.\footnote{We may -- without loss of generality -- assume that $H$ has non-negative eigenvalues.}
Writing out $H$ in its spectral decomposition $H = \sum_{i} \epsilon_{i} P_{i}$ with its eigenvalues $\epsilon_{i}$ and eigenprojections $P_{i}$, we see that the mere application of unitaries $U$ would only traverse the boundary of the variational cone as they map the projections $P_{i}$ again to projections $U^{\dag} P_{i} U$.
Meanwhile, $\malpha^{\dag} P_{i} \malpha^{\vphantom{\dag}}$ generally is not a projection, thus the application of $\malpha$ offers fundamentally new optimisation directions through the interior of the variational cone.

The optimisation problem can now be formulated as an $\ell$-dimensional generalised eigenvalue problem (GEP): $\mathbf{H} \bm{\alpha} = \lambda \mathbf{E} \bm{\alpha}$, by introducing the moment matrices $\mathbf{E}_{i j} = \braket{\phi | U_{i}^{\dag} U_{j}| \phi}$ and $\mathbf{H}_{i j} = \braket{\phi | U_{i}^{\dag} H U_{j}| \phi}$.
In the \hyperref[subsection:DerivationOfTheGeneralisedEigenvalueProblem]{Appendix}, we provide an elegant derivation from \eqref{equation:ParameterOptimisation} which contrasts the approaches in \cite{McClean2017HybridQuantumClassicalHierarchyForMitigationOfDecoherenceAndDeterminationOfExictedStates,Huggins2020ANonOrthogonalVariationalQuantumEigensolver}.
The correct normalisation is ensured by the additional constraint that $\bm{\alpha}^{\dag} \mathbf{E} \bm{\alpha} = 1$.
The matrix elements of $\mathbf{E}$ and $\mathbf{H}$ can be efficiently estimated via, e.g., the Hadamard test \cite{Kitaev1995QuantumMeasurementsAndTheAbelianStabilizerProblem} or the protocol introduced in \cite{Binkowski2024OneForAllUniversalQuantumConicProgrammingFrameworkForHardConstrainedCombinatorialOptimizationProblems}.

In contrast to the complex and hard \cite{Bittel2021TrainingVariationalQuantumAlgorithmsIsNPHard} optimisation problems which arise for the parameter optimisation of unitary PQCs, we can track which influence shot noise when recording the moment matrices $\mathbf{E}$ and $\mathbf{H}$ has on the GEP's solution.
Namely, we find that solving a perturbed GEP which arises from taking $m$ samples for the measurement of each moment matrix entry, respectively, yields with probability $1 - \varepsilon$ a solution in an $\mathcal{O}((m \log \varepsilon)^{-1/2})$ neighbourhood around the solution of the ``perfect'' GEP without shot noise.
More details can be found in the \hyperref[subsection:RobustnessUnderShotNoise]{Appendix}.

Another key point here is that $\ell$ may be chosen independently of $n$ and small enough that standard classical routines can quickly solve the GEP -- for the QAOA$_{p}$, the size of the auxiliary GEP is given by $\ell = 3$.

The update $\ket{\phi} \mapsto \ket{\phi'}$ according to \eqref{equation:AnsatzClass} can be achieved via a nondeterministic LCU channel:
As derived in the \hyperref[subsection:LCUChannelImplementation]{Appendix}, preparing some state $\ket{\psi(\bm{\alpha})}_{\ancilla}$ in a $\lceil \log_{2}(\ell)\rceil$-qubit ancilla register $\anchil$, applying the unitary gate $\mathcal{U} \equiv \sum_{i = 1}^{\ell} U_{i} \otimes \ketbra{i}{i}_{\ancilla}$ to the product state $\ket{\phi} \ket{\psi(\bm{\alpha})}_{\ancilla}$, performing a projective measurement on some other state $\ket{\xi(\bm{\alpha})}_{\ancilla}$ in the ancilla system, and then postselecting on the measurement's success can be fine-tuned to yield $\ket{\phi'}$ on the main register with a success probability of
\begin{align}\label{equation:MethodSuccessProbability}
    p_{\text{succ}} = \norm{\bm{\alpha}}_{1}^{-2}.
\end{align}

\section{Results}\label{section:Results}

\begin{figure*}[!th]
    \begin{minipage}[t]{0.5\linewidth}
        \includegraphics[scale=0.51]{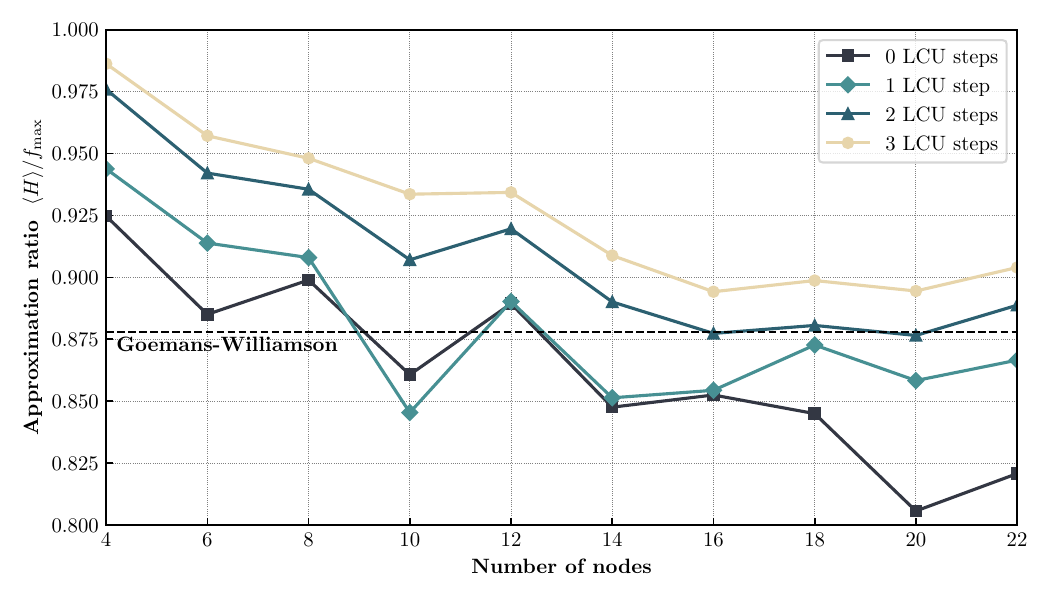}
    \end{minipage}\hfill
    \begin{minipage}[t]{0.5\linewidth}
        \includegraphics[scale=0.51]{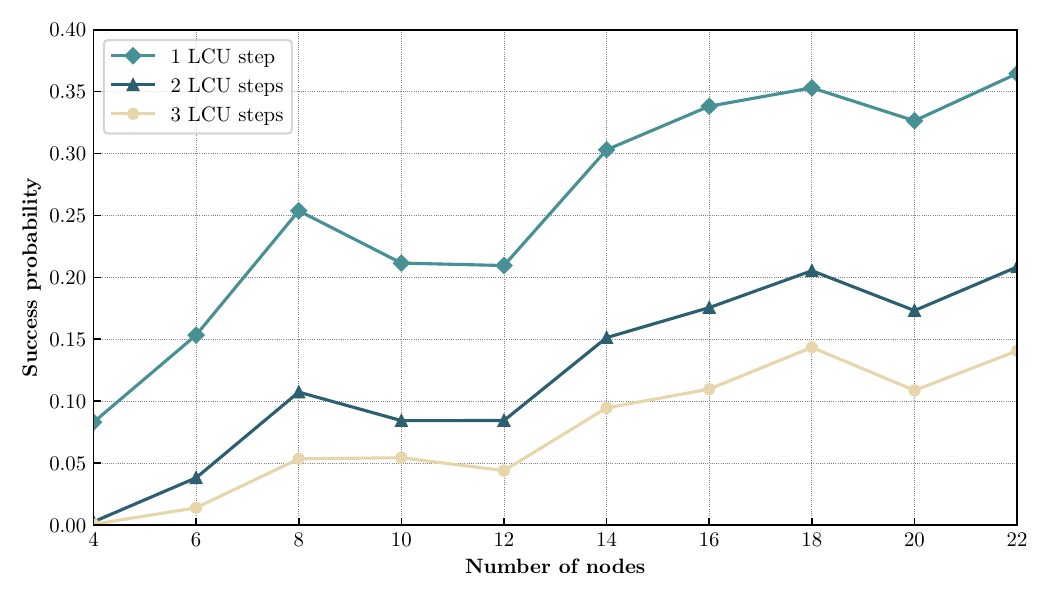}
    \end{minipage}
\caption{
    Performance of QAOA with and without LCU assistance across various MAXCUT instances.
    \textbf{a)} We simulate QAOA using gradient descent as the classical optimisation routine, with and without the inclusion of LCU steps, on 3-regular graphs.
    The black curve represents the baseline QAOA without LCU assistance, which fails to surpass the Goemans-Williamson (GW) bound for graphs with 14 or more nodes.
    The green curve corresponds to QAOA with a single LCU step, which exhibits similar limitations for larger instances.
    The blue curve shows results for QAOA with two LCU steps, which consistently performs near or above the GW bound even for larger problem sizes.
    Finally, the beige curve denotes QAOA assisted by three LCU steps, maintaining approximation ratios well above the GW bound across all tested instance sizes.
    All four methods show a general decline in performance as problem size increases.
    \textbf{b)} Success probabilities for the LCU-assisted QAOA variants, shown using the same colour scheme as in a).
    The LCU-free implementation is omitted as it succeeds deterministically.
    Interestingly, success probabilities tend to increase with instance size, contrary to the expectation that larger instances are inherently more difficult.
    While this trend may eventually reverse for sufficiently large problems, the observed stability in the moderate-size regime suggests that high-quality solutions can be obtained reliably with a modest number of repetitions.
}
\label{figure:MultiResults}
\end{figure*}

We demonstrate the capabilities of our LCU-assisted QAOA in noise-free simulations on a four-qubit to 22-qubit instances of MAXCUT.
The underlying graphs are randomly drawn from the class of three-regular graphs and their edges are equipped with random integer weights between one and three.

We employ a simple yet effective variant of the QAOA whose optimisation routine iteratively utilises the objective value's gradient at the current point in parameter space for a gradient descent, starting with $p = 2$.
Whenever the descent of the subsequent iteration falls short of a predefined threshold, meaning the algorithm creeps on a barren plateaus or is caught in a local minimum, we apply the described LCU step.

At each point of stagnation, a new computational branch is initiated: for each prior QAOA branch, we take its final quantum state as the initial state for a new QAOA$_4$.
This recursive branching framework provides a natural baseline to evaluate the effectiveness of the QCP intervention relative to another layer of QAOA.

We begin by analysing the results for the largest problem instance in greater detail.
As shown in \hyperref[figure:Results]{Figure 3 \textbf{a})}, the application of QAOA without LCU assistance quickly stagnates after approximately ten iterations, yielding an approximation ratio of about $0.785$, substantially below the Goemans-Williamson (GW) bound of $0.878$ \cite{Goemans1995ImprovedApproximationAlgorithmsForMaximumCutAndSatisfiabilityProblemsUsingSemidefiniteProgramming}.
Introducing a single LCU step increases the approximation ratio to $0.842$.
A subsequent QAOA stage, initiated from this point, further improves the ratio to $0.856$ before also reaching stagnation.
Appending a second LCU step lifts the approximation ratio to $0.885$, thereby surpassing the GW bound.
While an additional round of QAOA fails to enhance the performance further, the introduction of a third LCU step raises the approximation ratio to $0.904$.

In parallel to each LCU step, we also investigate the effect of appending a two-layer QAOA circuit.
In the LCU-free case, the optimisation encounters a barren plateau, from which it only escapes after roughly 60 iterations, ultimately reaching a modest approximation ratio of $0.821$.
Two further appended QAOA layers fail to provide additional gains, suggesting convergence to a local optimum from which the purely unitary ansatz cannot escape.
Similarly, following the first LCU step with further QAOA layers, we observe that approximately 350 iterations are required to overcome a barren plateau, ultimately reaching an approximation ratio of around $0.867$. In contrast, the second LCU step yields an immediate improvement that crosses the GW bound.
After two LCU steps, further QAOA layers no longer improve performance, again suggesting a local optimum.
Nevertheless, a third LCU step still yields a significant enhancement in the approximation ratio.
These observations support the hypothesis that LCU steps are capable, in principle, of overcoming both barren plateaus and local optima.

The observed improvements in solution quality come at the cost of reduced success probability.
The first LCU step decreases the probability of success by approximately $63.5\%$ (as defined in \eqref{equation:MethodSuccessProbability}), while the probability of successfully applying three consecutive LCU steps drops to approximately $14.1\%$.

The quality of the resulting solutions becomes even more apparent when comparing the final output distributions of the raw QAOA and the triply LCU-assisted QAOA, as shown in \hyperref[figure:Results]{Figure 3 \textbf{b})}.
Without LCU assistance, the QAOA produces a state with minimal overlap with highly optimal solutions.
In contrast, after three LCU steps, the resulting state is significantly biased toward computational basis states associated with high-quality solutions, albeit with a slight increase in amplitudes corresponding to suboptimal states.

To assess scalability, we also perform the same numerical experiments for all even qubit counts between 4 and 20 (3-regular graphs do not exist for odd numbers of vertices).
This allows us to track both approximation ratios and success probabilities across increasing problem sizes.
As anticipated, and shown in \hyperref[figure:MultiResults]{Figure 4 \textbf{a})}, the approximation ratios generally decline with increasing instance size across all tested configurations of QAOA layers and LCU steps.

With the exception of the 10-qubit case, the QAOA circuit without LCU assistance consistently yields the lowest approximation ratios.
In contrast, performance improves progressively with the number of LCU steps applied.
Starting from 14 qubits, at least two LCU steps are required to exceed the Goemans-Williamson bound.

Interestingly, the success probabilities associated with the application of one, two, and three LCU steps do not diminish with problem size.
On the contrary, they exhibit a mild increase as the number of qubits grows.
While we do not expect this trend to persist for substantially larger instances -- given the NP-hard nature of the problem -- it is notable that, within the moderate regime considered in this work, high-quality approximation ratios can be achieved with probabilities exceeding $10\%$, even for the largest tested instances.

\section{Conclusion}\label{section:Conclusion}

In this paper we have proposed a general method of extending existing VQAs to include non-unitary ansatz classes.
We have justified, both conceptually and numerically, that these augmentations naturally avoid phenomena like barren plateaus and local traps.
Furthermore, we have introduced a concrete NISQ-compatible implementation recipe for updating the ansatz state on the main register.
More specifically, our method comprises a non-deterministic LCU channel for which we showed a reasonable success probability.
Open problems remain, such as an in-depth performance comparison with noisy qubits and further improvements for the success probability of the LCU channel.
Controlling the latter would then immediately lead to an iterative version of our LCU step that could render the underlying VQA obsolete.

\section*{Acknowledgment}

This work was supported by the DFG through SFB 1227 (DQ-mat), Quantum Frontiers, the Quantum Valley Lower Saxony, the BMBF projects ATIQ and QuBRA. Helpful correspondence and discussions with Mark Bennemann, Andreea-Iulia Lefterovici, Arne-Christian Voigt, Reinhard F.\ Werner, and Sören Wilkening are gratefully acknowledged.

\bibliographystyle{IEEEtran}
\bibliography{IEEEabrv.bib,bibliography.bib}

\appendix

\subsection{\label{subsection:DerivationOfTheGeneralisedEigenvalueProblem}Derivation of the Generalised Eigenvalue Problem}

We derive the emergence of a generalised eigenvalue problem for the parameter optimisation \eqref{equation:ParameterOptimisation}.
First and foremost, note that the introduced moment matrices $\mathbf{E}$ and $\mathbf{H}$ fulfil
\begin{align*}
    \bm{\alpha}^{\dagger} \mathbf{E} \bm{\alpha} = \sum_{i, j = 1}^{\ell} \overline{\alpha}_{i} \braket{\phi | U_{i}^{\dagger} U_{j}^{\vphantom{\dagger}} | \phi} \alpha_{j} = \braket{\phi | \malpha^{\dagger} \malpha^{\vphantom{\dagger}} | \phi}
\end{align*}
and
\begin{align*}
    \bm{\alpha}^{\dagger} \mathbf{H} \bm{\alpha} = \sum_{i, j = 1}^{\ell} \overline{\alpha}_{i} \braket{\phi | U_{i}^{\dagger} H U_{j}^{\vphantom{\dagger}} | \phi} \alpha_{j} = \braket{\phi | \malpha^{\dagger} H \malpha^{\vphantom{\dagger}} | \phi}.
\end{align*}
Hence, \eqref{equation:ParameterOptimisation} may be restated as
\begin{align*}
\begin{split}
    & \min_{\bm{\alpha} \in \C^{\ell}}  \bm{\alpha}^{\dagger} \mathbf{H} \bm{\alpha} \\
    & \text{s.t. } \bm{\alpha}^{\dagger} \mathbf{E} \bm{\alpha} = 1.
 \end{split}
\end{align*}
Its dual function is defined by \cite{Boyd2004ConvexOptimization} 
\begin{equation}\label{eq:dualfunc}
    g(\lambda) = \lambda + \inf_{\bm{\alpha} \in \C^{\ell}} \bm{\alpha}^{\dag} \left(\mathbf{H} - \lambda \mathbf{E} \right) \bm{\alpha}
\end{equation}
If $\mathbf{H} - \lambda \mathbf{E} \prec 0$, \eqref{eq:dualfunc} is unbounded below because there exists some $\Tilde{\bm{\alpha}} \in \C^{\ell}$ with $\Tilde{\bm{\alpha}}^{\dag} (\mathbf{H} - \lambda \mathbf{E}) \Tilde{\bm{\alpha}} < 0$ such that
\begin{equation*}
    g(\lambda) = \lambda + \lim_{r \to \infty} (r \Tilde{\bm{\alpha}})^{\dag} \left(\mathbf{H} - \lambda \mathbf{E} \right) r \Tilde{\bm{\alpha}} = -\infty.
\end{equation*}

Since $\mathbf{E}$ and $\mathbf{H}$ are bounded and hermitian the feasible set of $\lambda$'s will be bounded and closed.
Hence we have that the optimal parameter $\lambda_{\opt}$ will be on the boundary of this feasible set and therefore yield an operator $\mathbf{H} - \lambda_{\opt} \mathbf{E}$ that has at least one zero eigenvalue, i.e., we have that there is a vector $\phi \in \C^{\ell}$ with
\begin{align}\label{equation:GEP}
     \left(\mathbf{H} - \lambda_{\opt} \mathbf{E} \right) \phi = 0 .
\end{align}
Finding such tuples $(\phi, \lambda_{\opt})$ is precisely a generalised eigenvalue problem.
We can now take
\begin{align*}
    \bm{\alpha}_{\opt} = \frac{\phi}{\sqrt{\phi^{\dagger} \mathbf{E} \phi}}.
\end{align*}
For this vector we have that 
\begin{align*}
    \bm{\alpha}_{\opt}^{\dagger} \mathbf{E} \bm{\alpha}_{\opt}^{\vphantom{\dagger}} = \frac{\phi^{\dagger} \mathbf{E} \phi}{\phi^{\dagger} \mathbf{E} \phi} = 1
\end{align*}
and, by using \eqref{equation:GEP},
\begin{align*}
     \bm{\alpha}_{\opt}^{\dagger} \mathbf{H} \bm{\alpha}_{\opt}^{\vphantom{\dagger}} 
    = \lambda_{\opt} \frac{\phi^{\dagger} \mathbf{E} \phi}{\phi^{\dagger} \mathbf{E} \phi}= \lambda_{\opt}
\end{align*}
which shows that $\bm{\alpha}_{\opt}$ is indeed the optimiser of \eqref{equation:ParameterOptimisation}.
 
\subsection{\label{subsection:LCUChannelImplementation}LCU channel implementation}

Given an optimal parameter vector $\bm{\alpha}$ stemming from the GEP, we now discuss possible implementations of the accordingly parameterised non-unitary gate \eqref{equation:AnsatzClass}.
As a main tool we assume a $\lceil \log_{2}(\ell)\rceil$-qubit ancilla register $\hil_{\ancilla}$ where we encode some information about $\bm{\alpha}_{\opt}$.

The first place to store such information and to introduce the unitaries $U_{i}$ is within an isometry $V : \hil \rightarrow \hil \otimes \hil_{\ancilla}$, embedding the main register in the system composed of main and ancilla register.
Given an arbitrary, but fixed state $\ket{\psi}_{\ancilla} \in \hil_{\ancilla}$, we consider $V$ to be of the form
\begin{align}
    V \coloneqq \sum_{i = 1}^{\ell} U_{i} \otimes \psi_{i} \ket{i}_{\ancilla}.
\end{align}
In practice, one achieves such an isometry by preparing the state $\ket{\psi}_{\ancilla}$ in the ancilla register and applying the unitary
\begin{align*}
    \mathcal{U} \coloneqq \sum_{i = 1}^{\ell} U_{i} \otimes \ketbra{i}{i}_{\ancilla}
\end{align*}
to the composite system.

Furthermore, we allow for scanning the ancilla register for some state $\ket{\xi}_{\ancilla} \in \hil_{\ancilla}$ while doing nothing in the main register.
This yields a map $P : \hil \otimes \hil_{\ancilla} \rightarrow \hil \otimes \C \cong \hil$ given by
\begin{align}
    P \coloneqq \one \otimes \bra{\xi}_{\ancilla}.
\end{align}
Together, $V$ and $P$ yield the following operation
\begin{align}
    PV = \sum_{i = 1}^{\ell} \psi_{i} \overline{\xi}_{i} U_{i}.
\end{align}
Choosing $\psi_{i} \overline{\xi}_{i} = \eta \alpha_{i}$ for every $i = 1, \ldots, \ell$, where the prefactor $\eta \in \R_{+}$ ensure correct normalisation, yields the requested non-unitary gate's action.

Note that $P$ -- and thus $P V$ -- is not norm-preserving.
The element of measurement involved in $P$ renders it nondeterministic, introducing a certain failure probability to not obtain the output state that corresponds to postselecting on the measurement's success.
In other words, for any state $\ket{\phi} \in \hil$, $P V \ket{\phi}$ is merely subnormalised.
Its norm square directly gives rise to the probability of obtaining success in the involved measurement, hence of implementing the correct output state:
\begin{align}\label{equation:SuccessProbability}
    p_{\text{succ}}
    &= \braket{\phi | V^{\dagger} P^{\dagger} P V | \phi} \nonumber \\
    &= \sum_{i, j = 1}^{\ell} \overline{\psi}_{i} \xi_{i} \psi_{j} \overline{\xi}_{j} \braket{\phi | U^{\dagger}_{i} U_{j} | \phi} \nonumber \\
    &= \eta^{2} \sum_{i, j = 1}^{\ell} \overline{\alpha}_{i} \mathbf{E}_{i j} \alpha_{j} \nonumber \\
    &= \eta^{2} \bm{\alpha}^{\dagger} \mathbf{E} \bm{\alpha} \nonumber \\
    &= \eta^{2}.
\end{align}

The success probability therefore only depends on the normalisation factor $\eta$.
The latter, in turn, can be influenced by the choice of $\ket{\psi}_{\ancilla}$ and $\ket{\xi}_{\ancilla}$.
The approach leading to \eqref{equation:MethodSuccessProbability} is given by the following setup:
Let $\sqrt{\bm{\alpha}}$ denote the vector obtained by component-wisely taking the principal square root of $\bm{\alpha}$.
Assigning
\begin{align}\label{equation:RootEncoding}
    \ket{\psi}_{\ancilla} \equiv \frac{\sqrt{\bm{\alpha}}}{\sqrt{\norm{\bm{\alpha}}_{1}}},\quad \ket{\xi}_{\ancilla} \equiv \frac{\overline{\sqrt{\bm{\alpha}}}}{\sqrt{\norm{\bm{\alpha}}_{1}}}
\end{align}
yields correctly normalised quantum states which fulfil $\psi_{i} \overline{\xi}_{i} = \alpha_{i} /  \norm{\bm{\alpha}}_{1}$ for all $i = 1, \ldots, \ell$.
Due to \eqref{equation:SuccessProbability}, this indeed yields \eqref{equation:MethodSuccessProbability}.

Another possible setup is
\begin{align}\label{equation:NaiveEncoding}
    \ket{\psi}_{\ancilla} \equiv \bm{\alpha} / \norm{\bm{\alpha}}_{2},\quad \ket{\xi}_{\ancilla} = \ket{+}_{\ancilla}
\end{align}
which, by \eqref{equation:SuccessProbability}, results in a success probability of
\begin{align}\label{equation:NaiveEncodingSuccessProbability}
    p_{\text{succ}} = \ell^{-1} \norm{\bm{\alpha}}_{2}^{-2}.
\end{align}
The advantage of this approach is that it concentrates the difficult part of the state preparation entirely on $\ket{\psi}_{\ancilla}$ while the state $\ket{\xi}_{\ancilla}$ would be cheap to prepare.
However, this approach yields a generally lower success probability than the prior method.

\subsection{\label{subsection:RobustnessUnderShotNoise}Robustness under shot noise}

In realistic scenarios, there will be some statistical error when estimating a pair of moment matrices $\mathbf{E}$ and $\mathbf{H}$ from finite-sized measurement data.
This error can, however, be easily analysed.
Due to linearity, the propagation of this error to the final cost function estimate can be controlled.
This is we can outline a robustness result that can be roughly paraphrased as:
When the sampling error is small, the propagated error to the cost function will also be small. 

In principle, there is a whole variety of methods for tomographing  moment matrices.
One of the most straight forward ways is to impose selfadjointness and directly measure the observables corresponding to the real and imaginary parts of the entries.\footnote{See \cite[Protocol I]{Binkowski2024OneForAllUniversalQuantumConicProgrammingFrameworkForHardConstrainedCombinatorialOptimizationProblems} for an improved protocol which reuses measurement results and relies on classical post-processing to calculate moment matrix entries.}
For simplicity, we will assume uniform sample sizes $m$ for the estimation of each entry.
Collecting this data and computing the mean will give us estimators $\Tilde{\mathbf{E}}$ and $\Tilde{\mathbf{H}}$ for our moment matrices.
A standard Hoeffding bound \cite{Vershynin2018HighdimensionalProbabilityAnIntroductionWithApplicationsInDataScience} can be applied in this situation, yielding a statement of the form:
With high probability $1 - \varepsilon$ the errors of these estimates are bounded, in Frobenius norm, by
\begin{align}
    \Vert\mathbf{E} - \Tilde{\mathbf{E}}\Vert_{F} &\leq C_{E} \delta(\varepsilon, m) \label{equation:HoeffdingE} \\
    \Vert\mathbf{H} - \Tilde{\mathbf{H}}\Vert_{F} &\leq C_{H} \delta(\varepsilon, m), \label{equation:HoeffdingH}
\end{align}
with constants $C_{E}$ and $C_{H}$ that depend on the spectrum of our target Hamiltonian $H$, and a scaling factor 
\begin{align}
    \delta(\varepsilon, m) = \sqrt{\frac{2 \ell^{2} (\log{2 \ell^{2}} -\log{\varepsilon})}{m}}.
\end{align}
We would now like to employ Lipschitz continuity bounds for GEPs \cite{Stewart1990MatrixPertubationTheory} in order to relate solution $\tilde{\bm{\alpha}}$ of the noisy GEP $\Tilde{\mathbf{H}} \tilde{\bm{\alpha}} = \lambda \Tilde{\mathbf{E}} \tilde{\bm{\alpha}}$ to the solution $\bm{\alpha}$ of the original GEP.

A subtle problem is, however, that the resulting GEP $\Tilde{\mathbf{H}} \tilde{\bm{\alpha}} = \lambda \Tilde{\mathbf{E}} \tilde{\bm{\alpha}}$ does not need to be feasible, even though the original GEP is, by design, always guaranteed to be feasible.
In order to ensure feasibility of the perturbed GEP, we have to make one more step and project our estimators onto the set of GEP-feasible matrices.
Given concrete estimates $\Tilde{\mathbf{E}}$ and $\Tilde{\mathbf{H}}$, we can find a good pair of feasible matrices $\overline{\mathbf{E}}$, $\overline{\mathbf{H}}$ by solving the optimisation
\begin{align*}
   \delta_{\text{SDP}} \coloneqq & \inf_{\overline{\mathbf{E}}, \overline{\mathbf{H}} \in \C^{\ell \times \ell}} \Vert\Tilde{\mathbf{E}} - \overline{\mathbf{E}}\Vert_{F} + \Vert\Tilde{\mathbf{H}} - \overline{\mathbf{H}}\Vert_{F} \\ 
    & \quad \ \, \text{s.t. } \quad \operatorname{supp}(\overline{\mathbf{H}})\subset \operatorname{supp}(\overline{\mathbf{E}}) \\
    & \quad \ \, \hphantom{\text{s.t. }} \quad \ \overline{\mathbf{E}} \in \text{Pos}_{\ell \times \ell}, \ \overline{\mathbf{H}} \in \text{Herm}_{\ell \times \ell}  
\end{align*}
which can indeed be formulated as a semidefinite program and solved at a similar time scale as the GEP.
After finding some concrete optimisers $\overline{\mathbf{E}}$ and $\overline{\mathbf{H}}$ out of the optimisation above, we can solve the GEP $\overline{\mathbf{H}} \overline{\bm{\alpha}} = \overline{\lambda} \overline{\mathbf{E}} \overline{\bm{\alpha}}$.
The distance of its solution $\overline{\bm{\alpha}}$ to the solution of the GEP without sampling errors can now be bounded by combining the Hoeffding bounds \eqref{equation:HoeffdingE} and \eqref{equation:HoeffdingH}, the GEP's Lipschitz continuity, as well as the fact that
\begin{align*}
    \Vert\mathbf{E} - \overline{\mathbf{E}}\Vert_{F} \leq \Vert \Tilde{\mathbf{E}} - \mathbf{E}\Vert_{F} \text{ and } \Vert\mathbf{H} - \overline{\mathbf{H}}\Vert_{F} \leq \Vert\Tilde{\mathbf{H}} - \mathbf{H}\Vert_{F}.
\end{align*}
In detail, we have
\begin{align}
    \Vert\bm{\alpha} - \overline{\bm{\alpha}}\Vert_{2} \leq C_{\text{GEP}} \delta(\varepsilon, m),
\end{align}
where the precise prefactor $C_{\text{GEP}}$ depends on the spectrum of $\mathbf{H}$, the unitaries in the LCU step, and also on the previous coefficients $C_{E}$ and $C_{H}$.
In order to judge the error introduced by measuring a Hamiltonian on states produced by an LCU step, it makes actually more sense to compute the trace distance between the projections onto $\overline{\bm{\alpha}}$ and $\bm{\alpha}$.
It is straight forward to check that this error is bounded by 
\begin{align}
    \Vert \ketbra{\bm{\alpha}}{\bm{\alpha}} -  \ketbra{\overline{\bm{\alpha}}}{\overline{\bm{\alpha}}} \Vert_{1} \leq \sqrt{2} C_{\text{GEP}} \delta(\varepsilon, m). 
\end{align}
In conclusion, we can state that the error in the value of the cost function introduced by finite sampling and shot noise will show the typical behaviour known form statistical concentration: the error will shrink with $1 /\sqrt{m}$. 
This result has to be seen in contrast to the error robustness of other parameterised quantum circuits, where an corresponding analysis of the error introduced by shot-noise can, to the best of our knowledge, not be bounded in a similar, straightforward manner, although we acknowledge some recent theoretical work regarding the influence of shot noise on the performance of variational quantum algorithms \cite{Gu2021AdaptiveShotAllocationForFastConvergenceInVariationalQuantumAlgorithms,Liu2025StochasticNoiseCanBeHelpfulForVariationalQuantumAlgorithms}.

\end{document}